\documentclass[12pt,letterpaper]{article}          
\usepackage{osajnl}
\usepackage[draft]{hyperref} 

\def\hwp{HWP}
\def\ahwp{AHWP}

\def\beq{\begin{equation}}
\def\eeq{\end{equation}}

\begin{document}

\bibliographystyle{unsrt}

\title{A Millimeter-Wave Achromatic Half Wave Plate}
\author{Shaul Hanany}
\address{School of Physics and Astronomy\\
  University of Minnesota\\
  116 Church St. SE, Minneapolis, Minnesota, 55455}
\email{hanany@physics.umn.edu}
\author{Johannes Hubmayr}
\address{School of Physics and Astronomy\\
  University of Minnesota\\
  116 Church St. SE, Minneapolis, Minnesota, 55455}
\author{Bradley R.\ Johnson}
\address{School of Physics and Astronomy \\
  University of Minnesota \\
  116 Church St. SE, Minneapolis, Minnesota, 55455 \\
  Present Address: School of Physics and Astronomy \\
  Cardiff University, \\
  5 The Parade, Queens Buildings \\
  Cardiff, CF24 3YB, UK}
\author{Tomotake Matsumura}
\address{School of Physics and Astronomy\\
  University of Minnesota\\
  116 Church St. SE, Minneapolis, Minnesota, 55455}
\author{Paul Oxley}
\address{School of Physics and Astronomy\\
  University of Minnesota\\
  116 Church St. SE, Minneapolis, Minnesota, 55455}
\newpage
\author{Matthew Thibodeau}
\address{School of Physics and Astronomy\\
  University of Minnesota\\
  116 Church St. SE, Minneapolis, Minnesota, 55455}

\begin{abstract}
We have constructed an achromatic half wave plate (AHWP) suitable for
the millimeter wavelength band. The \ahwp\ was made from a stack of
three sapphire a-cut birefringent plates with the optical axes of the
middle plate rotated by 50.5~degrees with respect to the aligned axes
of the other plates.  The measured modulation efficiency of the
\ahwp\ at 110~GHz was $96 \pm 1.5$\%. In contrast, the modulation
efficiency of a single sapphire plate of the same thickness was $43
\pm 4$\%. Both results are in close agreement with theoretical
predictions. The modulation efficiency of the \ahwp\ was constant as
a function of incidence angles between 0 and 15 degrees.  We discuss
design parameters of an \ahwp\ in the context of astrophysical broad
band polarimetry at the millimeter wavelength band.
\end{abstract}

\ocis{120.5410, 230.5440, 260.5430, 350.1260.}

\maketitle 

\section{Introduction}
\label{sec:introduction}

Half wave plate (\hwp) retarders are used extensively for polarimetric
measurements. The technique is used across a broad range of
electro-magnetic frequencies because it provides an effective way to
discriminate against systematic errors.  The modulation efficiency of
a \hwp\ that is constructed from a single birefringent plate can reach
100\% for a set of discrete electro-magnetic frequencies but away from
these frequencies the efficiency drops rapidly.  To overcome this
limitation it has been proposed to stack several birefringent plates
with specific relative angles of their optical
axes~\cite{pancharatnam55,title81,title75}.  Such a construction has
been called achromatic \hwp\ (\ahwp) because it has a broader
frequency range over which the polarimetric efficiency is high
compared to a \hwp\ that is made from a single plate.  The efficiency
of an \ahwp\ depends on the number of plates in the stack and on their
relative angles.  The concept has been demonstrated experimentally in
the optical and IR bands~\cite{tinbergenbook}.  Murray
et~al.~\cite{murray97} have described briefly measurements of an
\ahwp\ made of 5 quartz plates for wavelengths between 350 and 850
$\mu$m and an \ahwp\ made of 3 quartz plates for wavelengths between 1
and 2~mm. No detailed information is given about the measurements, the
analysis, tests for systematic errors, or about the optimization of
the \ahwp\ in respect to the relative angles between the plates.  A
2-element achromatic waveguide polarizers for operation at $\sim$1~cm
is mentioned by Leitch et al.~\cite{leitch02b}.

In this paper we present the construction of a sapphire \ahwp\ and
measurements of its properties at a wavelength of 2.7~mm (110~GHz).
We also present an analysis of the design of a three-plate sapphire
\ahwp\ for a wavelength of 2~mm. There is currently interest in 
an \ahwp\ that is suitable for the mm-wave band because of the
increase in experimental efforts to measure the polarization of the
cosmic microwave background radiation. Several experiments will use
\hwp s and increasing the bandwidth where the efficiency is high will
increase the signal-to-noise ratio of the experiment.

\section{Experimental Setup}

\begin{figure}
\centerline{\rotatebox{0}
{\scalebox{1.0}{\includegraphics{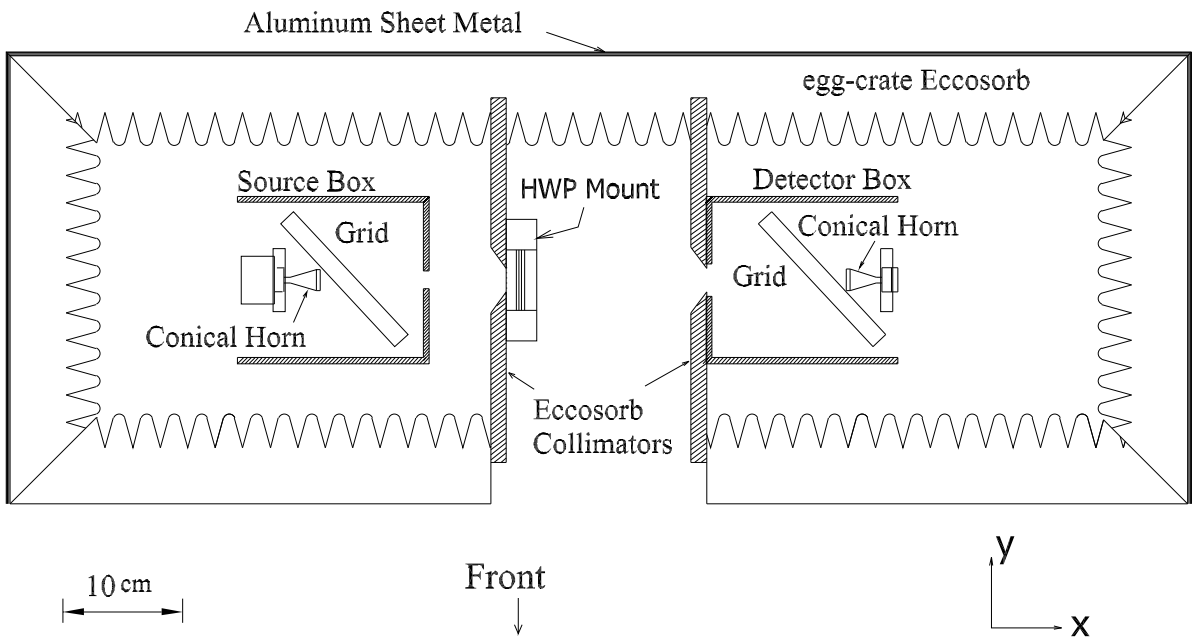} } } }
\caption{A top view sketch of the experimental setup.}
\label{fig:exp_setup}
\end{figure}

A top-view sketch of the experimental setup is shown in
Figure~\ref{fig:exp_setup}.  We used a Gunn
oscillator~\cite{spacek_source} at 110~GHz and a diode
detector~\cite{spacek_detector} as a source and detector of radiation,
respectively. Both source and detector had conical horns
that provided beams of 12~degrees full width at
half maximum. They emitted and were sensitive to linearly polarized
radiation with a -15~dB maximum level of cross polarization at
$\sim$10 degrees from peak gain. The source and detector were aligned
by maximizing the signal received by the detector as a function of its
orientation relative to the fixed orientation of the source.

We used wire grid polarizers to increase the level of linear
polarization of the light emitted by the source and detected by the
detector. The grids, which were made by Buckbee-Meers, were measured
to have a modulation efficiency of 97\%~\cite{johnson_thesis}.

The source, detector, and polarizers were housed in metallic boxes
that were lined inside (outside) with Emerson and Cuming Eccosorb
LS-16 (LS-14). One side of the boxes was open.

Between the boxes were placed two 1.25~cm thick plates of Emerson and
Cuming Eccosorb MF-124 which served as collimators. They had
19~mm diameter knife-edged holes which faced the source. The knife
edges were covered with 0.07~mm thick aluminum tape.  The \hwp\ was
installed between the collimators in a 5~cm diameter Newport mount
that could rotate around the $x$ axis with a resolution of
1~degree. The mount was held by a cylindrical leg that gave it another
degree of freedom for rotation around the $z$ axis.  The beam filled
the central 15\% area of the \hwp. Its angular extent when it reached
the detector was 2~degrees.

The entire experiment was mounted on a metallic optics bench.
Aluminum sheet metal lined with egg-crate Eccosorb-CV3
enclosed the experiment from three sides. Egg-crate Eccosorb was also
placed both in front and above the source and detector boxes, as 
shown in Figure~\ref{fig:exp_setup}.

\section{Achromatic Half Wave Plate}

We used a stack of three sapphire a-cut plates to construct the
\ahwp. Each of the fine ground plates had a thickness of
$2.32\pm0.05$~mm, which made each a \hwp\ for a frequency of 193~GHz.
The three plates were mounted together with a front and back
anti-reflection coating made of 0.35~mm thick polished Herasil.  The
orientation of the second plate was rotated by 50.5~deg with respect
to the orientation of the aligned first and third plates.  We had an
angular accuracy of $\pm 1$~degree in assembling the stack and an
accuracy of $\pm 1.5$~degrees in orienting the stack-mount normal to
the incoming beam.  The ordinary and extraordinary axes of any of the
plates were known to within 0.5~degree.

We compared the performance of the \ahwp\ to the performance of a
`chromatic' plate, a single a-cut plate of sapphire with a thickness
of 2.32~mm. The chromatic plate was stacked with the same layers of
anti-reflection coating as the \ahwp.

We used a frequency of 110~GHz to make the measurements because at
this frequency the difference between the modulation efficiency of
the \ahwp\ and of the single plate are nearly maximized
thereby providing a clear demonstration
of the achromaticity of the stack.

\section{Measurements, Analysis, and Results} 

\begin{figure}
\centerline{\rotatebox{-90}
{\scalebox{.7}{\includegraphics{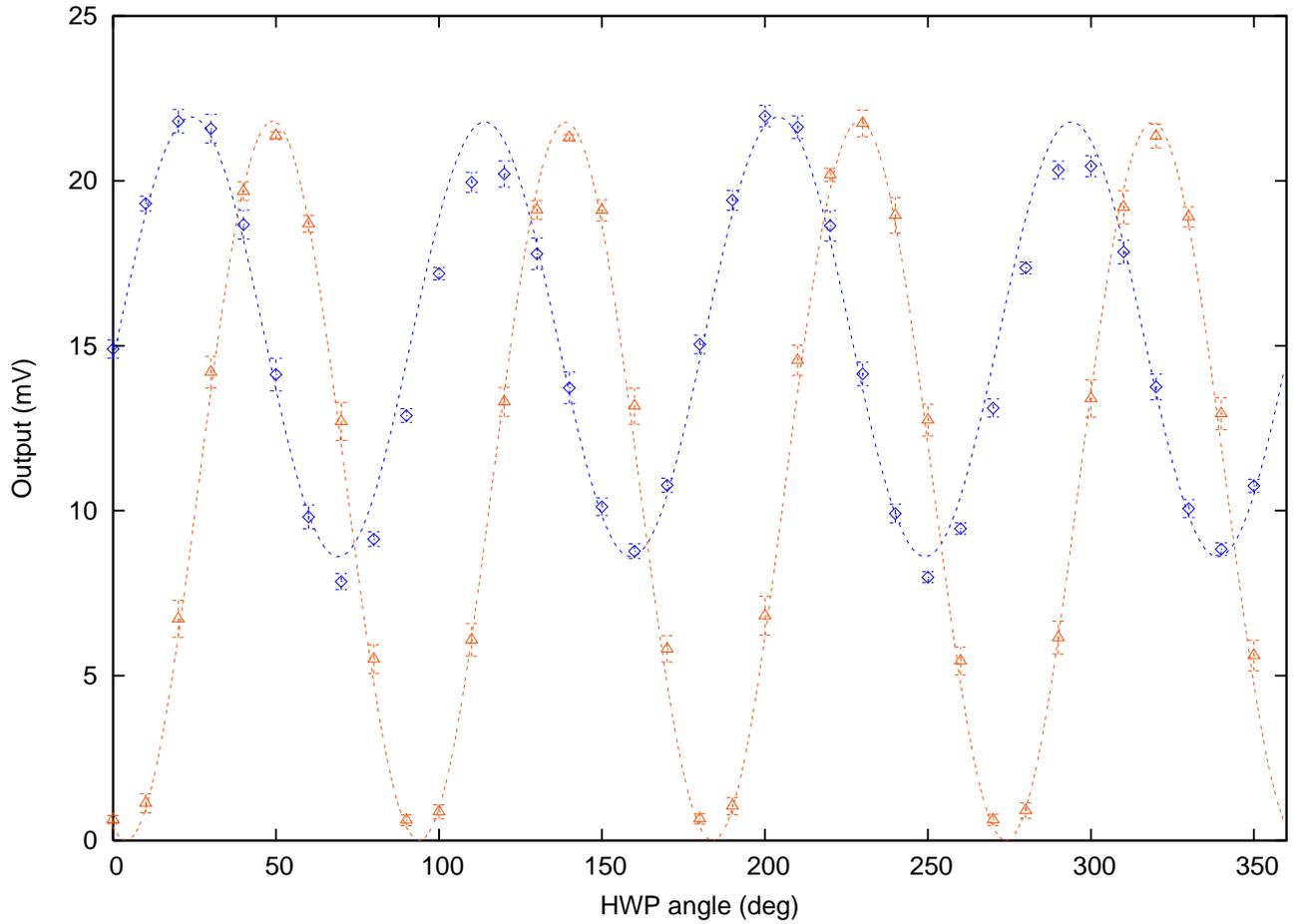}}}}
\caption{Measurements (points) and theoretical
predictions (dash) of the signal detected as a function of
rotation angle of the plates for the chromatic plate (blue diamond)
and for the \ahwp\ (red triangles). Error bars are the standard
deviations of 5 repeated measurements. The theoretical predictions
have no free parameters.}
\label{fig:data}
\end{figure}

To quantify the efficiency of the plates we measured the detected
intensity as a function of their rotation angle $\alpha$ about the $x$
axis. Data were taken every 10~degree in angle and are shown in
Figure~\ref{fig:data}.  Error bars are the standard deviation of 5
repeat measurements of the efficiency.  A repeat measurement consisted
of assembling all individual pieces into a stack, mounting the stack,
and taking data. No changes in other elements in the experiment were
made between repeat measurements.

A constant offset of about 0.7~mV was measured when the aperture of
the detector box was blocked and was subtracted from the data. This
level was constant with rotation of the plates, between different
independent measurements of a given stack, and between measurements
with different stacks. The data was then fit with the following
model
\beq 
D = \sum_{i=0}^{8} A_{i}\cos(i\alpha + \phi_{i}).  
\label{eqn:hwp_model}
\eeq
The output of the fitting were the 9 amplitudes and 8 phases, where
$\phi_{0}$ was set to zero. The modulation efficiency was defined as
\beq
\epsilon = { A_{4} \over A_{0} }. 
\label{eqn:efficiency}
\eeq 
The value of $\epsilon$ did not change when we fit the data only up to
the fourth harmonic (5 amplitudes and 4 phases). The quality of the
fit however degraded from a reduced $\chi^2$ of 0.27 and 0.9 for the
achromatic and chromatic plates, respectively, with 8 harmonics to 5.8
and 2.6, respectively, with 4 harmonics.

Predictions about the efficiency of the plates were calculated using
the technique of Mueller matrices. The intensity of the light incident
on the detector was generated by multiplying an incident Stokes vector
representing 100\% $Q$ polarized light by Mueller matrices that
simulated the response of the two anti-reflection layers, the plates,
and a 100\% $Q$ polarized detector. An overall normalization was taken
from a measurement of the power detected in the absence of a \hwp\ in
the light path. The phase was taken from the known orientation of the
plates. Normal incidence was assumed throughout. A prediction for the
detected intensity was calculated as a function of $\alpha$ in steps
of 1~degree, fitted by the model given in
Equation~\ref{eqn:hwp_model}, and a predicted efficiency was
calculated using Equation~\ref{eqn:efficiency}. The predicted response
of the plates as a function of angle is shown in
Figure~\ref{fig:data}. The prediction shown is not a fit to the
data. There are no free parameters in this prediction.

\begin{figure}
\centerline{\rotatebox{90}
{\scalebox{0.7}{\includegraphics{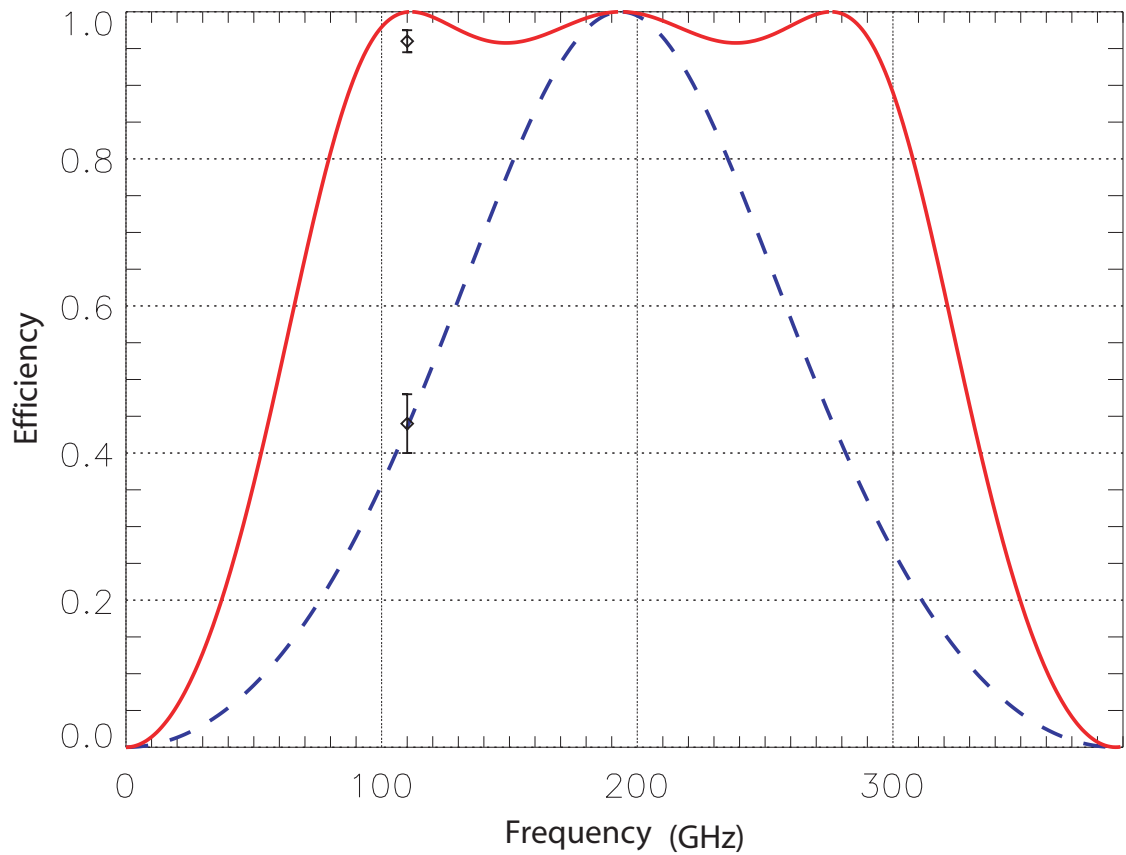}}}}
\caption{The predicted modulation 
efficiency as a function of frequency of the \ahwp\ (red broad) and of
a single plate (blue narrow) and the measured
efficiency of both plates.}
\label{fig:efficiency}
\end{figure}

Figure~\ref{fig:efficiency} shows the predicted efficiency of the
chromatic and achromatic plates as a function of frequency and our
measured values of $43 \pm 4 \%$ and $96 \pm 1.5 \%$, respectively.
The predicted values are 43.5\% and 100\%, respectively.  Uncertainty
in the predicted values of the efficiency, due to uncertainty
in the indices of sapphire~\cite{lamb96}, is 1.5\% for the single plate
and negligible for the \ahwp. The errors on the measurements of
the modulation efficiency were calculated by summing the statistical 
and an estimate of the systematic errors in quadrature. 

We also measured the efficiency for angles of incidence that are not
normal by tilting the \ahwp\ about the $z$ axis between angles of zero
and 15 degrees. We found no change in the efficiency as a function of
angle within statistical errors.


Spurious signals generated by reflections can be a source of
systematic errors. We checked the level of signal detected by the
detector when either of the collimators were blocked with metal or
with a piece of Eccosorb MF124. The level was 0.7 mV for all cases and
did not change as a function of the rotation angle of the plates in
their mount.

The experiment was repeated for various distances of the plates from
the source. The efficiency of the single plate varied in a sinusoidal
manner with position with an amplitude of 1.9 \% and a period of 1.4
mm.  This period is also half the wavelength of the source and we
hypothesize that reflections in the setup cause the small variation in
efficiency. We have also observed that the shape of the deviations
between the theoretical prediction for the detected signal and the one
measured vary as a function of the position of the plate. The data
shown in Figure~\ref{fig:data} is representative of the magnitude of
such deviations.  For the achromatic plate the peak-to-peak changes in
efficiency as a function of distance were smaller than the quoted
statistical error.

\section{Discussion} 

There is good agreement between each of the no-free-parameters
predictions shown in Fig.~\ref{fig:data} and the data. Both the
predicted overall modulation amplitude and the relative phase shift
are reproduced by the measurements. The measured modulation
efficiencies are close to the predicted values. 

\begin{figure}
\centerline{\rotatebox{90}
{\scalebox{.6}{\includegraphics{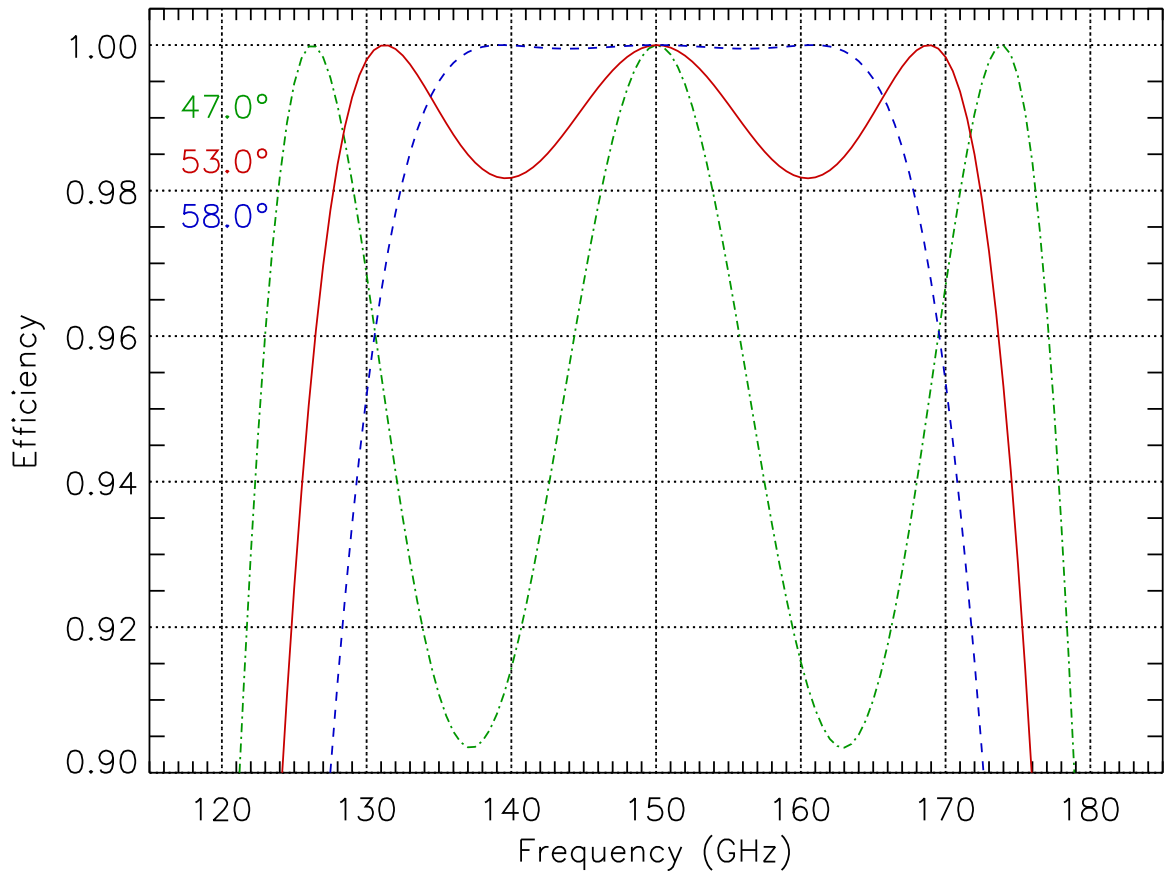}}}}
\caption{Predicted modulation 
efficiency of an \ahwp\ as a function of frequency near 150~GHz for
rotation angles of 47 (dash dot, green), 53 (solid, red) and 58 (dash,
blue)~degrees of the second plate. Each sapphire plate in the stack is
a \hwp\ for a frequency of 50~GHz.}
\label{fig:breadth_ripple}
\end{figure}

\begin{figure}
\centerline{\rotatebox{90}
{\scalebox{.9}{\includegraphics{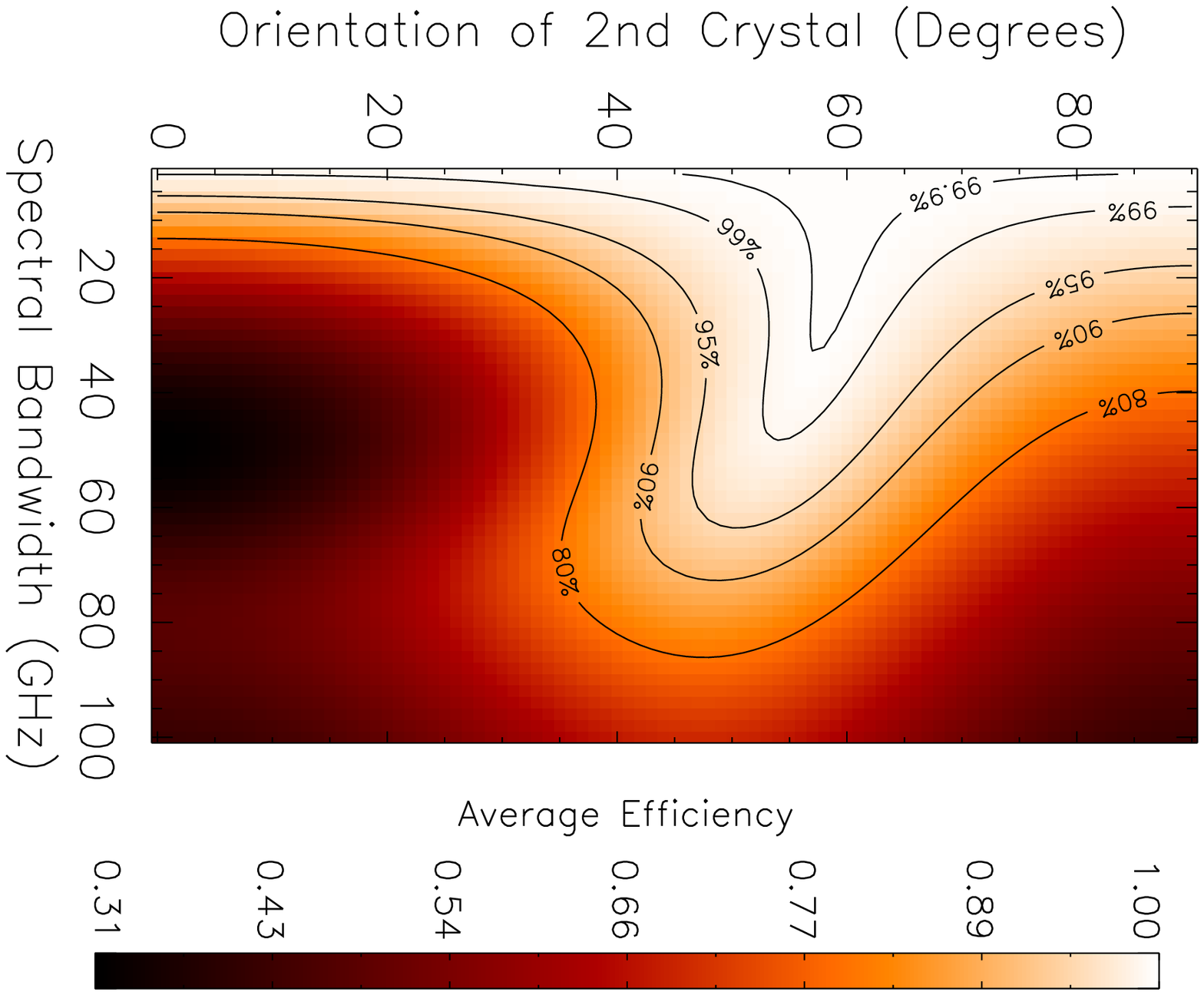}}}}
\caption{The average modulation efficiency
(color scale and contours) as a function of the orientation of the
second plate and the spectral width of a top hat band centered on
150~GHz (for example, a width of 60~GHz means $150 \pm
30$~GHz).}
\label{fig:contour}
\end{figure}

\ahwp's can be constructed with various combinations of birefringent
plates each giving a different degree of
achromaticity. Title~\cite{title75} showed that with 3 plates of the
same material an \ahwp\ should have the first and last plates aligned
and most of our discussion is restricted to such a
stack. Figure~\ref{fig:breadth_ripple} shows the efficiency of an
\ahwp\ made of three sapphire plates as a function of frequency and
for three different orientations of the second plate. Each of the
plates is a \hwp\ at odd harmonics of 50~GHz, suitable for a 
cosmic microwave background polarization experiment - EBEX - that 
we are currently constructing. EBEX will operate at 150, 250, 350 and
450~GHz. An orientation angle of 58~degrees gives close to a constant
modulation efficiency over a band of $\sim$40~GHz.  A plate
orientation angle of 47~degrees gives a band of $\sim$60~GHz at the
expense of variations of the efficiency within that band.  It is
therefore interesting to quantify the {\it average} modulation
efficiency as a function of bandwidth and as a function of rotation
angle of the second plate. The results are shown in
Figure~\ref{fig:contour} for a top-hat frequency response and they
demonstrate several features. The maximum average efficiency decreases
as a function of bandwidth but with a proper choice of angle average
efficiencies that are larger than 95\% are achievable with up to
60~GHz of bandwidth. The angular precision required for the
orientation of the second plate is rather coarse. The efficiency for
60~GHz of bandwidth is larger than 95\% for any angle between 47 and
56~degrees. Even smaller accuracy is required for narrower bandwidths.

\begin{figure}
\centerline{\rotatebox{0}
{\scalebox{1.0}{\includegraphics{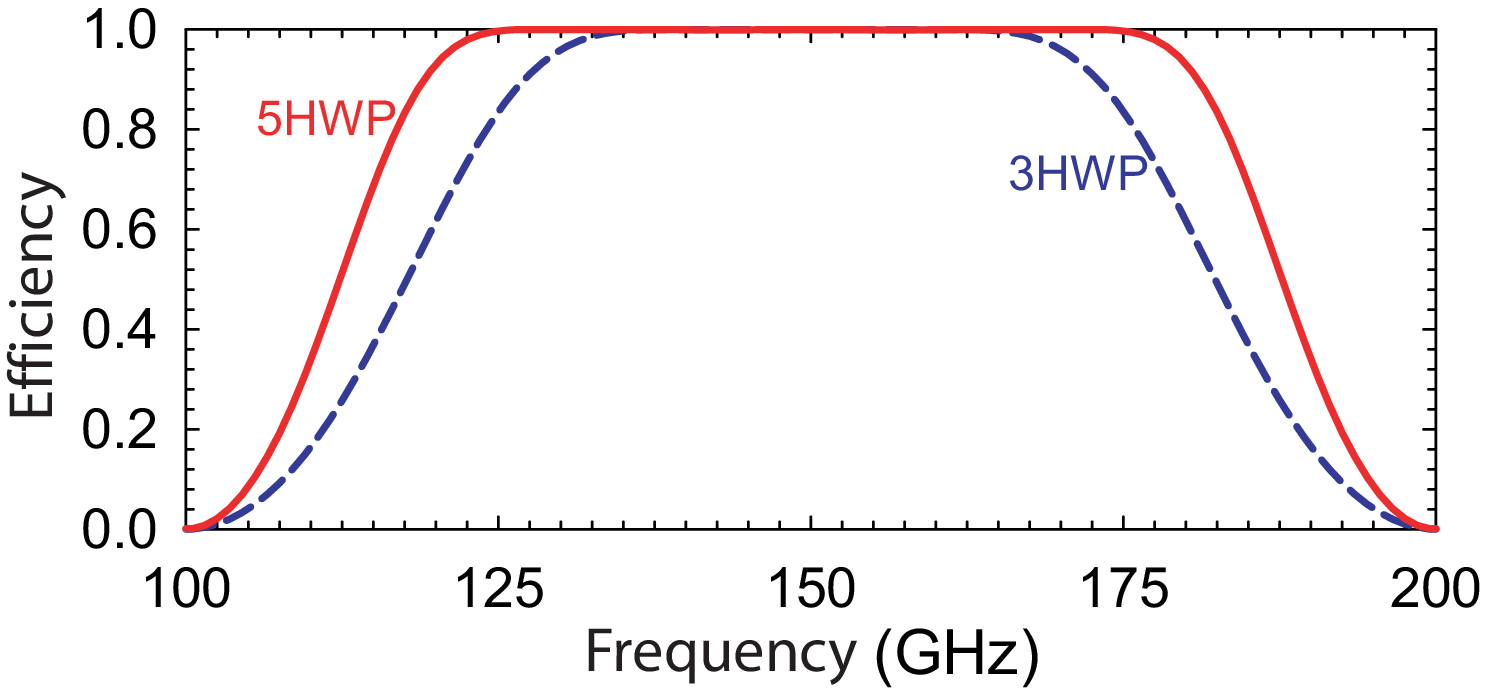}}}}
\caption{The modulation efficiency of an
\ahwp\ made of a stack of 5 plates compared to the modulation
efficiency of an \ahwp\ made of a 3-stack. The 5-stack has orientation
angles of 28.8, 94.5, 28.8 and 2~degrees for the plates after the
first, respectively.  For the 3-stack the second plate is at
57.5~degrees. Each of the plates is sapphire and is optimized for
50~GHz.}
\label{fig:5stack}
\end{figure}

\begin{figure}
\centerline{\rotatebox{90}
{\scalebox{1.0}{\includegraphics{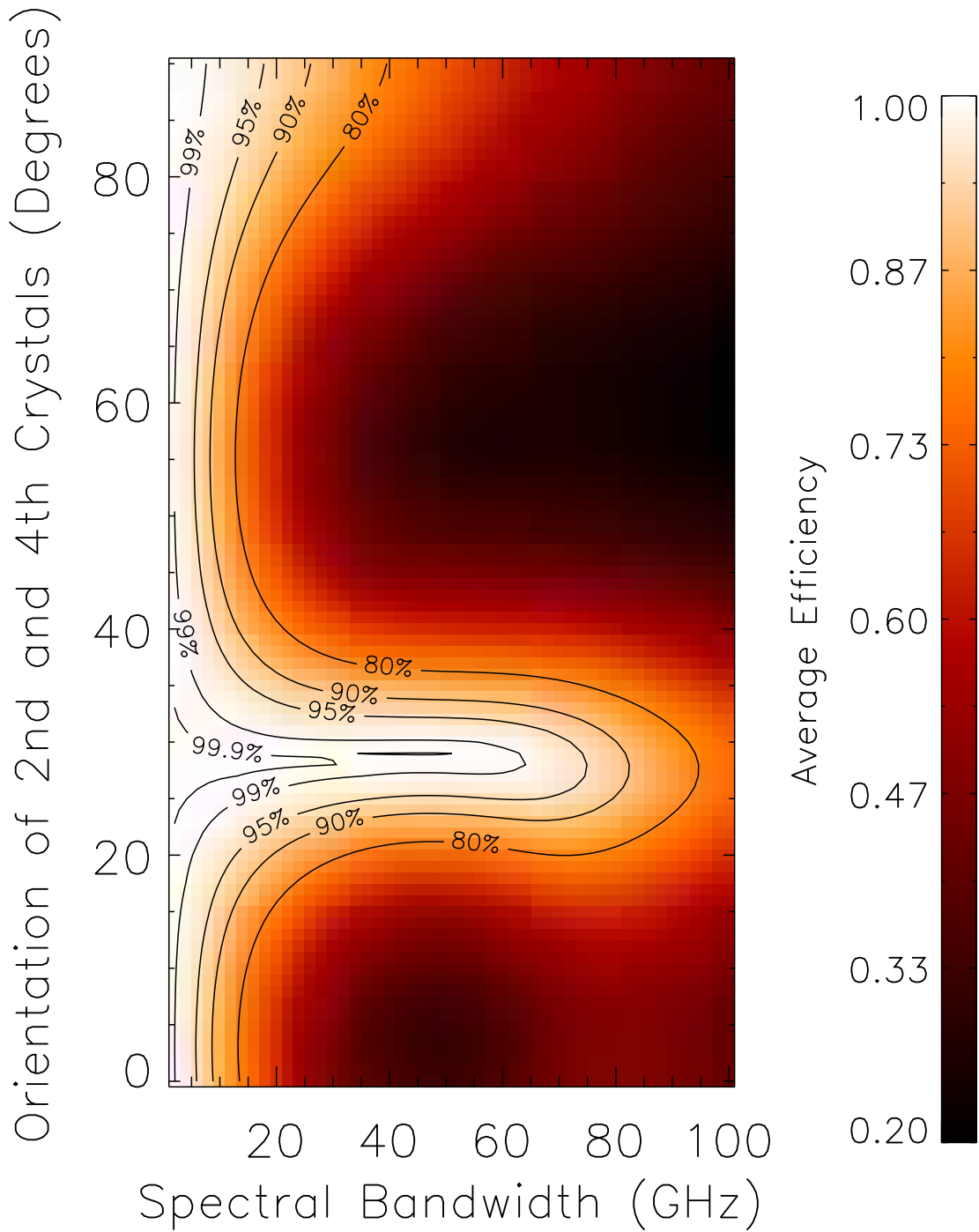}}}}
\caption{The average modulation
efficiency (color scale and contours) for an \ahwp\ made of a
5-stack. The efficiency is given as a function of the orientation of
the second and fourth plates (relative to the first) and the spectral
width of a top hat band centered on 150~GHz. The relative angles of
the third and fifth plates are 94.5 and 2~degrees, respectively.}
\label{fig:5stack_int}
\end{figure}

A stack of 5 plates can give high modulation efficiency over an even
broader range of frequencies compared to a 3-stack; see
Figures~\ref{fig:5stack} and~\ref{fig:5stack_int}.  With an assumption
of a top-hat frequency response of the instrument we calculate that
for the balloon-borne EBEX the penalty in increased absorption and
emission from the thicker stack of sapphire plates would be smaller
than the increase in signal and therefore a properly designed 5-stack
would increase the signal-to-noise ratio of the experiment.

Interest in mm-wave \ahwp\ has increased recently because of the
scientific interest in the polarization of the cosmic microwave
background radiation. Several experiments including our own EBEX
are proposing to use \hwp's as means to modulate the
incident polarization~\cite{oxley04,church03}. The results presented
in this paper provide reassurance that these experiments can rely on
an \ahwp\ and that the efficiency of such a plate is constant for 
a relatively broad range of incidence angles. 

\end{document}